\documentclass[10pt,letter,twocolumn]{article}

\usepackage{amssymb,amsmath}

\usepackage{graphicx}

\begin{document}

\title{\bf Dynamics of soap bubble bursting and its implications to volcano acoustics}
\author{V. Vidal$^{1,2}$, M. Ripepe$^3$, T. Divoux$^1$, D. Legrand$^{4,5}$,
J.-C. G\'eminard$^{1,2}$, and F.~Melo$^1$ \\
\small $^1$Laboratorio de F\'isica No Lineal and CIMAT, Departamento de F\'isica, \\
\small Universidad de Santiago de Chile, Avenida Ecuador 3493, Santiago, Chile. \\
\small $^2$Laboratoire de Physique, Universit\'e de Lyon, Ecole Normale Sup\'erieure - CNRS, \\
\small  46 All\'ee d'Italie, 69364 Lyon Cedex 07, France. \\
\small $^3$Dipartimento di Scienze della Terra, Universit\`a degli Studi di Firenze, \\
\small  via La Pira, 4 - 50121 Firenze, Italy. \\
\small $^4$Universidad de Chile, Departamento de Geof\'isica, Blanco Encalada 2002, Santiago, Chile. \\
\small $^5$Now at: Universidad Aut\'onoma de M\'exico, Ciudad Universitaria, Instituto \\
\small  de Geof\'isica, Departamento de Vulcanolog\'ia, Del. Coyoacan, CP 04510, M\'exico DF.
}
\maketitle

%
%

\begin{abstract}
In order to assess the physical mechanisms at stake when giant gas bubbles burst
at the top of a magma conduit, laboratory experiments have been performed. 
An overpressurized gas cavity is initially closed by a thin liquid film, which suddenly bursts.
The acoustic signal produced by the bursting is investigated.
The key result is that the amplitude and energy of the acoustic signal 
strongly depend on the film rupture time. As the rupture time is uncontrolled in 
the experiments and in the field, the measurement of the acoustic excess pressure in the
atmosphere, alone, cannot provide any information on the overpressure inside the
bubble before explosion. This could explain the low energy partitioning between infrasound, 
seismic and explosive dynamics often observed on volcanoes.
\end{abstract}

%
%

\section{Introduction}

Volcanic explosions generate both seismic and acoustic waves propagating in the ground
and in the atmosphere, respectively. Monitoring the acoustic emissions thus represents, 
together with the seismic signals monitoring, an attractive tool to investigate the
source of volcanic explosions. In particular, the 
simultaneous recording of the seismic and acoustic signals might provide clues
to constrain the source process (e.g. \cite{Vergniolle96a}).
However, the link between the seismic and acoustic waves and the explosive source 
dynamics is still poorly understood. Some authors claim that seismic and acoustic waves are 
generated by an unique shallow process ($<500$~m depth) \cite{Kobayashi05,Johnson07}.
Others propose that the 
acoustic waves are produced by the bursting of meter-sized gas bubbles, while the seismic 
waves result from the pressure variations, in the magma column, associated with the rise 
of the gas bubbles toward the surface \cite{Ripepe01,Chouet03,James04}. Nonetheless,
almost all studies assert that acoustic waves are generated either by the bursting 
of the gas bubbles \cite{Ripepe96,Johnson03a} or by the oscillation 
of the magma membrane covering the gas slug just before the bursting \cite{Vergniolle96a}. 

The acoustic wave characteristics, in the infrasonic range, are commonly related to the 
properties of the bursting bubble, such as its volume and overpressure before bursting
\cite{Vergniolle96a}. However, most of these analysis are theoretical and numerical 
(e.g. \cite{Vergniolle96a}). Only a few laboratory experiments were 
dedicated to characterizing the acoustics of bubble bursting in conditions that are 
relevant to volcanology \cite{James04,James09}. 

Here we investigate experimentally the bursting, in static
conditions, of a 'slug' whose parameters (geometry and overpressure) are accurately 
controlled. The characteristics of the acoustic signal emitted at bursting (frequency, energy) 
are compared with the initial bubble geometry (volume) and overpressure.
This experiment focuses on the physical mechanisms at stake when the overpressurized 
cavity suddenly opens. Because the dynamics of bubble bursting on volcanoes is much 
more complex, we will not compare directly our experiment with the field situation. 
However, the physical processes we describe here are likely to be involved when a large gas
bubble explodes at the top of a volcano vent (Figure~1a). We therefore comment 
the results in regard to potential implications for large bubbles bursting on volcanoes. 
We discuss further (section~5) the limits of this application.

\section{Experimental results}

Our experimental setup consists of a cylindrical cavity drilled in a plexiglas slab (Figure~1b). 
Following \cite{Vidal06}, we close the cavity by stretching a thin soap film.
Air is injected inside and, due to the increase in the inner pressure, the thin soap film deforms 
and bulges out. Injection is stopped when a chosen overpressure $\Delta P$ is reached.
The system then remains in mechanical equilibrium, while the soap film drains the liquid aside 
\cite{Mysels59} and eventually bursts. 
This controlled experiment makes possible to easily vary the length and
volume of the cavity, and the gas overpressure before the film bursting. 
Different tube lengths $L$ (from 2 to 23~cm) and diameters $d$  (6, 8 or 10~mm) have 
been used (aspect ratio $\alpha=L/d$ ranging from 2 to 23) in order to quantify the role 
of the conduit geometry. 

The rupture of the film results in a sudden drop of the inner overpressure, which excites 
resonant modes inside the cavity.
The inner standing waves are damped due to dissipation along the walls and radiation out of
the open end of the cavity. A microphone inside the tube (Figure~1b, bottom) records the
pressure variation at the cavity bottom ($P_{int}$) while a microphone outside 
(Figure~1b, top) monitors the radiated acoustic waves ($P_{ext}$).
As expected for a resonating tube, both signals, inside and outside, exhibit the same 
spectral content, with a fundamental frequency associated with the wavelength in air 
$\lambda_0 \sim 4L$ and odd harmonics \cite{Kinsler82}. Note that the location of the 
inner microphone is pertinent, as the amplitude of the pressure variation associated
with all the harmonics is maximum at the cavity bottom.

\begin{figure}[t]
\begin{center}
\noindent\includegraphics[width=1\columnwidth]{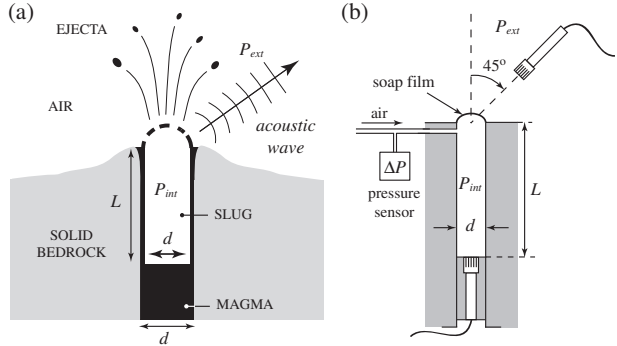}
\end{center}
\caption{\small
(a) Sketch of a slug exploding at the top of a volcanic conduit. 
b) Experimental setup of a soap film bursting at the surface of a cavity of well controlled 
geometry (length $L$, diameter $d$, volume $V$) and initial overpressure ($\Delta P$).}
\end{figure}

\section{Partitioning of the acoustic pressure}
\label{sec:partitioning}

\begin{figure}[h]
\begin{center}
\noindent\includegraphics[width=0.9\columnwidth]{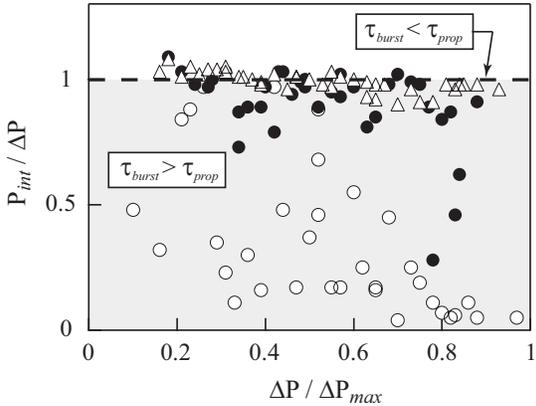}
\end{center}
\caption{\small
Normalized amplitude $P_{int}/\Delta P$ of the acoustic signal at bursting, inside the cavity, 
as a function of the initial normalized overpressure $\Delta P/\Delta P_{max}$.  
$P_{int}/\Delta P<1$ (gray region) indicates a slow dynamics of the film opening.
[Symbol,$\alpha$]: [$\circ$,$2$]; [$\bullet$,$8$]; [$\triangle$,$23$].}
\end{figure}
When the soap film breaks at the top of the cavity, the overpressure recorded by the 
bottom microphone (Figure~1b) drops from $+\Delta P$ to $-P_{int}$. Before the 
bursting, the overpressure $\Delta P$ inside the cavity is constant. We thus expect, in the 
absence of significant energy loss, to measure $P_{int}=\Delta P$. We found that this is 
true only for long tubes ($\alpha=23$, large gas volume) whereas for short tubes 
($\alpha=2$, small gas volume) a large scatter of the pressure drop is observed, 
with $P_{int} \leq \Delta P$ (Figure 2). 

This scatter can be explained by taking into account the film rupture dynamics, and 
in particular its typical rupture time $\tau_{burst}$ \cite{Vidal06,Divoux08}. 
It is approximated, from the experimental data, as the time necessary for the 
overpressure at the cavity bottom to drop from $+\Delta P$ to $-P_{int}$. 
The characteristic film rupture time is compared to the propagation time 
$\tau_{prop}$ of the acoustic wave inside the tube, defined as 
\begin{equation}
\tau_{prop}=2L/c
\end{equation}
where $c$ is the sound velocity ($c=340$~m/s). 
The experiment indicates that the initial relative amplitude $P_{int}/\Delta P$ is 
a strongly decreasing function of the ratio $\tau_{burst}/ \tau_{prop}$ (Figure~3a).
\begin{figure}[htb]
\begin{center}
\noindent\includegraphics[width=0.9\columnwidth]{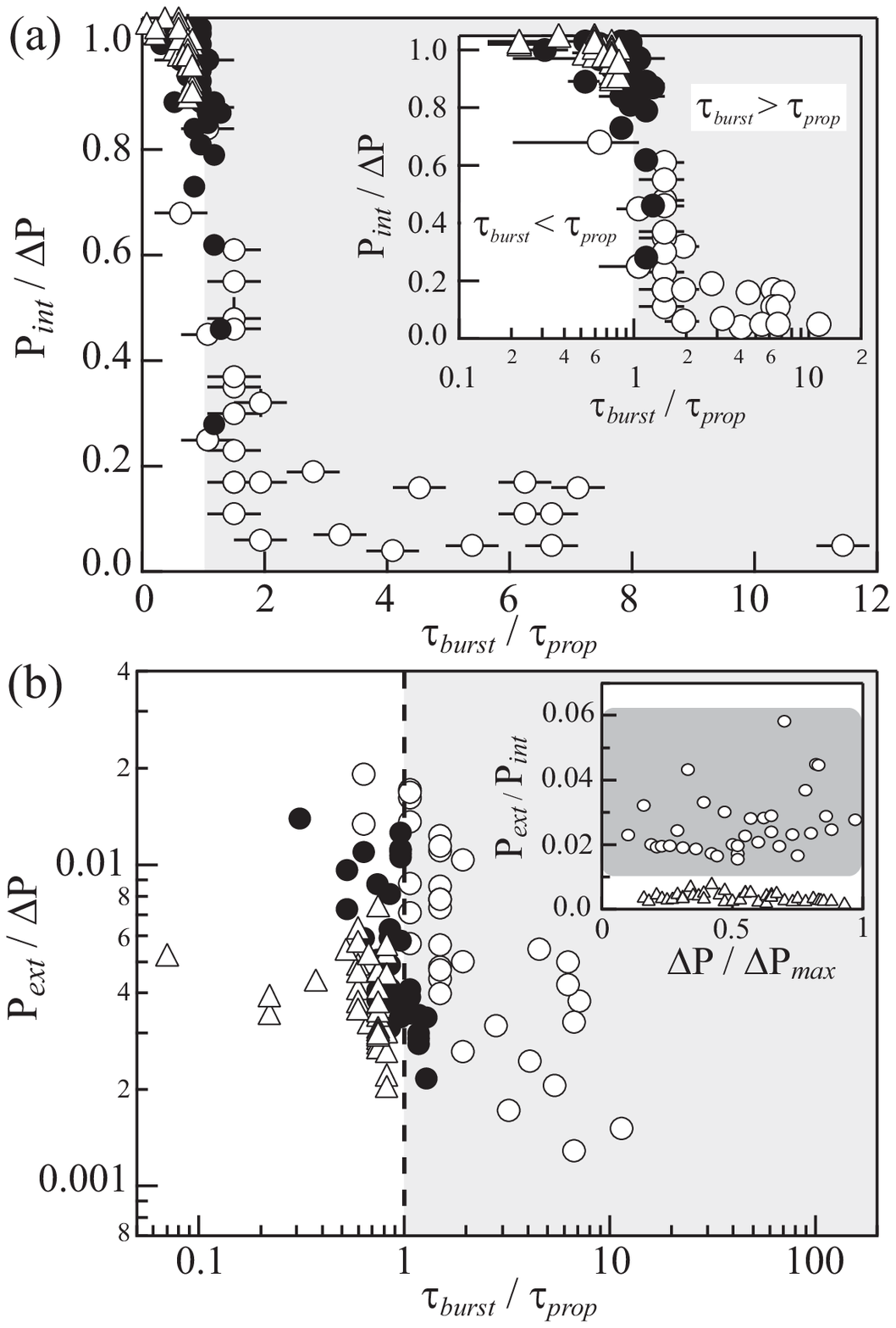}
\end{center}
\caption{\small
Normalized amplitude of the acoustic signal at bursting, inside (a) and outside (b) the cavity,
as a function of the ratio between the bursting time $\tau_{burst}$, and the propagation time
$\tau_{prop}$. $P_{ext}$ is recorded at $r=5$~cm from the cavity aperture.
The light gray region indicates a slow rupture dynamics of the film ($\tau_{burst}/\tau_{prop}>1$,
same as Figure~2). {\it Insets:} (a) Semi-log plot of $P_{int}/\Delta P$ as a function of 
$\tau_{burst}/\tau_{prop}$. The efficiency of the bubble bursting to transmit pressure waves 
in the air drastically drops for slow rupture dynamics ($\tau_{burst}/\tau_{prop}>1$). 
(b) Amplitude ratio $P_{ext}/P_{int}$. The most efficient energy partitioning occurs for 
the shorter tube (dark gray region). 
[Symbol,$\alpha$]: [$\circ$,$2$]; [$\bullet$,$8$]; [$\triangle$,$23$].}
\end{figure}
In other words, for a given geometry, the larger the characteristic rupture time is, the 
smaller is the amount of energy transferred to the resonant modes \cite{Vidal06}.
For $\tau_{burst}/\tau_{prop}>1$, we observe a drastic drop in the amplitude $P_{int}$
of the signal inside the cavity. 
Long cavities (large aspect ratio, e.g. $\alpha=23$) are not sensitive to the film
rupture time, as they always fulfill the condition $\tau_{burst}<\tau_{prop}$.
In this case, the acoustic amplitude inside the tube well approaches the initial overpressure, 
and $P_{int}/\Delta P \simeq 1$ (Figures~2 and 3a). To the contrary,
for short cavities, the system is sensitive to the film rupture time, which is on the order of
$\tau_{prop}$, and a large scatter is observed.
The uncontrolled bubble rupture time thus accounts for the scatter of the acoustic energy
measured outside the cavity ($P_{ext}$, see section~4).

On volcanoes, we do not have access to the acoustic waves inside the slug, and we
measure the acoustic signal propagating outside. When a slug bursts, 
part of the energy is radiated as infrasonic waves in the atmosphere, while part remains 
trapped in the volcanic body and propagates as seismic waves. We consider here that the 
overpressure recorded by the internal microphone ($P_{int}$, Figure~1b) represents the 
amplitude (and/or the energy) of the seismic signal generated by the explosive process, 
while the pressure outside the cavity ($P_{ext}$) is comparable to the amplitude of the 
acoustic waves recorded in the atmosphere as infrasonic waves.

Therefore, by comparing the amplitude ratio between the internal and external pressure variations
(Figure~3b, inset), we assume to observe a process similar to the amplitude (or energy) 
partitioning between seismic and acoustic waves associated with the same explosive dynamics.
Our experiment indicates that the inside and outside pressure partitioning ($P_{ext}/P_{int}$) 
changes as function of the tube length and hence, of the bubble volume. Longer tubes have 
acoustic waves with long propagation time and with large damping effects due to viscous 
dissipation. As a consequence, the amplitude of the acoustic signal outside the cavity will 
be much smaller than for short tubes (Figure~3b, inset).

In other words, long tubes (e.g. $L=23$~cm) are more efficient in terms of acoustic wave 
energy trapped inside the cavity (Figures~2 and 3a), but are not efficient to transmit acoustic 
energy outside (Figure~3b, inset). In contrast, short tubes (e.g. $L=2$~cm) 
are more efficient in radiating acoustic energy outside (Figure~3b, inset), but are also more 
sensitive to the rupture time ($\tau_{burst}/\tau_{prop}>1$, Figure~3a). If the energy loss 
for long tubes is mainly due to the non-efficient radiation process (small $P_{ext}/P_{int}$), 
the energy loss for short tubes (higher $P_{ext}/P_{int}$) is mainly due to the film rupture time 
($\tau_{burst}/\tau_{prop}>1$). In both cases, energy is dissipated.

\section{Energy balance}

In order to quantify the total energy balance in the system, we estimate the acoustic energy 
$E_a$ from the signal measured outside:
\begin{equation}
E_a = \frac{2 \pi r^2}{\rho c} \int_{t=0}^{\infty}P^2_{ext}(t) \, dt
\end{equation}
where $r$ is the distance between the cavity aperture and the microphone, $\rho$ the gas 
density, and $c$ the sound velocity in air. The potential energy stored inside the 'slug' 
before bursting can be written as:
\begin{equation}
E_p = \frac{1}{2} \frac{V \Delta P^2}{\rho c^2}
\end{equation}
where $V$ is the volume of the gas slug. Figure~4 displays the acoustic energy measured 
outside, as a function of the initial slug overpressure, before bursting. The three different 
experimental conditions represent the field situation, where the conduit radius is constant in time, 
but the slug length can vary from one explosion to another \cite{Vergniolle96a}. 
We report in Figure~4 the maximum total acoustic energy estimated from a series of
10 measurements, performed in similar conditions (same $\Delta P$). Each point then
represents the acoustic energy obtained when the rupture time is the smallest. Its effect
is therefore considered negligible ($\tau_{burst}/\tau_{prop}<1$).

For small initial overpressure $\Delta P$, 
the acoustic energy $E_a$ behaves as $E_a \sim \Delta P^2$. When $\Delta P$ increases, 
the bubble deforms and the soap film curvature increases. Consequently, when the film 
bursts, the pressure front entering the cavity is spherical, and thus no longer matches 
the planar geometry of the resonant modes. As a consequence, the efficiency of the 
energy transfer decreases and $E_a$ drops down when $\Delta P$ is increased (Figure~4).
Finally, we point out that this simple bubble bursting cannot build up gas 
overpressure above the maximum threshold. For a bubble bursting in static condition, 
the threshold pressure $\Delta P_{max}$ is given by  $\Delta P_{max}= 8 \sigma/d$
(semispherical film), where $\sigma$ represents the film surface tension.

In summary, two mechanisms limit the energy tranfer to the acoustic waves. 
First, the characteristic rupture 
time of the film breaking (section~3); Second, the film curvature before bursting. 
Indeed, as the film curvature increases, the pressure of the 
acoustic wave front generated at bursting departs from a planar wavefield. It is therefore 
less efficient in exciting the cavity resonant modes. This explains why, in the 
experiment, the fraction of energy, $E_a/E_p$, measured outside is small ($\sim$15\%) 
\cite{Vidal06}, and points out the importance of the geometry and dynamics of the film 
rupture in controlling the amount of energy released in the atmosphere as acoustic waves.

On volcanoes, the velocity of the film rupture and the film geometry itself are much more 
complex than in our experiments performed in static conditions, and depend largely on 
the dynamics of the bubble rising and on viscosity, temperature and volatile content of the 
magma film layer above the bubble.

\begin{figure}[b]
\begin{center}
\noindent\includegraphics[width=0.9\columnwidth]{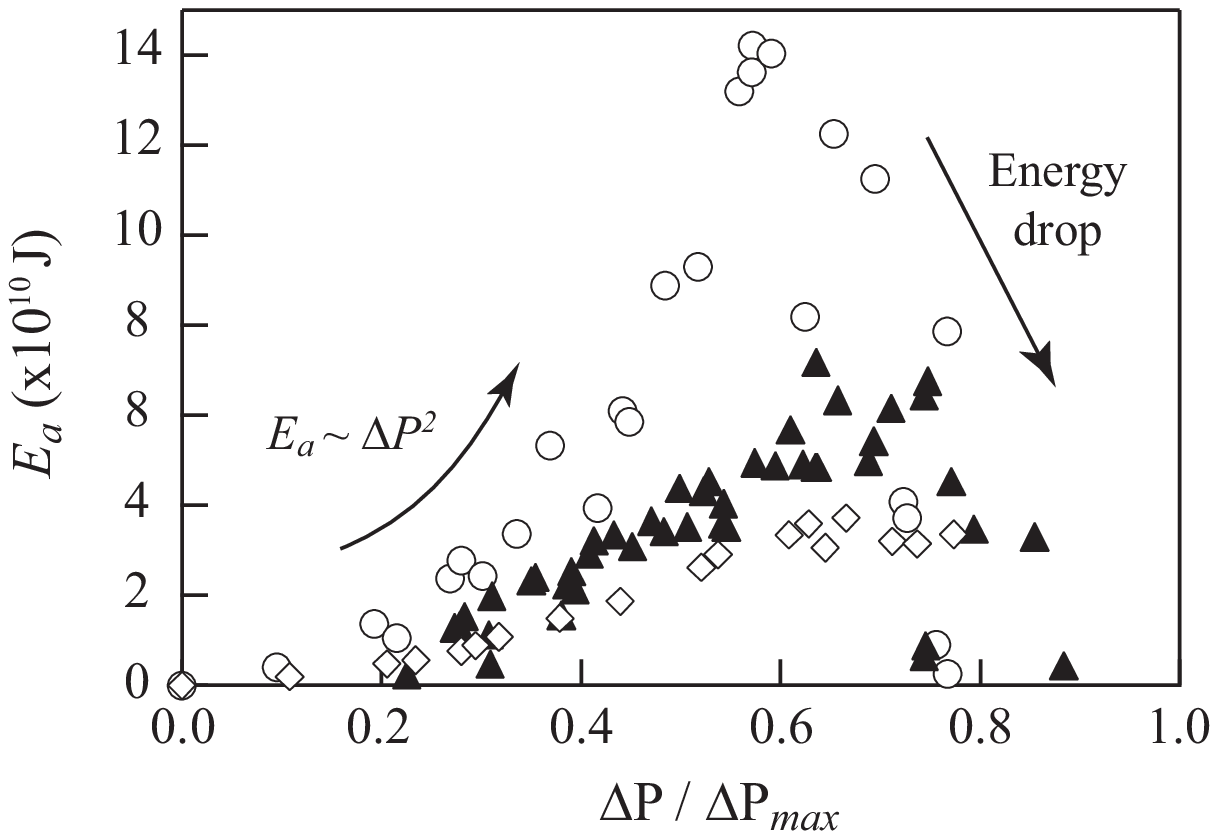}
\end{center}
\caption{\small
Acoustic energy $E_a$ as a function of the initial overpressure $\Delta P$ 
(normalized to $\Delta P_{max}$) for different lengths (same diameter $d=8$~mm). 
All data represent the maximum pressure recorded outside 
over a series of ten measurements. We assume that this value corresponds to the signal 
recorded when the rupture time is negligible ($\tau_{burst}/\tau_{prop}<1$). 
[Symbol,$\alpha$]: [\large$\circ$\normalsize,$2.5$]; [$\blacktriangle$,$5$];
[$\diamond$,$6.9$].}
\end{figure}

\section{Discussion and Conclusion}

This simple experiment provides an insight into the physical mechanisms involved 
in the bursting of a slug of well-controlled geometry and overpressure, in static conditions.
Even in a fully controlled laboratory experiment, the amplitude 
and energy of the pressure wave propagating into the 
atmosphere after bursting cannot be predicted from the initial slug overpressure - 
and vice versa. We demonstrated that two processes are responsible for this
unpredictability: (1) the rupture time of the bubble film, which cannot
be controlled in the experiments; and (2) the energy loss due to the film 
curvature at bursting, which excites more or less efficiently the cavity. 
When the rupture time $\tau_{burst}$ is larger than the 
characteristic propagation time $\tau_{prop}$ inside the cavity, the acoustic signal amplitude 
(and, thus, the energy) drops. The energy fraction ($E_a / E_p$) transferred into the 
acoustic signal radiated outside decreases drastically when the rupture time 
$\tau_{burst}$ increases. 

A quantitative comparison with the much more complex field situation is out of the scope 
of this paper. Indeed, when a slug bubble rises and bursts in a volcanic conduit, 
viscous and inertial forces - two processes not investigated in this work - 
play an important role. On the one hand, by limiting the bubble expansion when rising,
these forces are thought to be responsible for the large overpressure stored inside the slug
before bursting \cite{James08}. On the other hand, viscous effects may strongly
affect the dynamics of the film aperture \cite{Debregeas95}. 
Experiments investigating bubbles bursting in either 
a Newtonian \cite{James08} or non-Newtonian \cite{Divoux08} fluids 
pointed out the importance of the rising velocity and, more generally, of the bursting 
dynamics, on the acoustic wave amplitude.

However, even if in static conditions, the physical mechanisms described here are likely 
to be at stake in the field. 
Different rupture dynamics and film thicknesses largely affect the rupture time, and the
acoustic signal emitted at bursting. In particular, we point 
out that any interpretation of the measured acoustic amplitude, or energy, in terms of gas 
overpressure in the bubble before bursting requires a good knowledge of the physics 
controlling the opening of the bubble at bursting. 
This suggests that on volcanoes also, the amplitude of the acoustic waves 
generated by the bubble bursting strongly depends on the thickness and on the rupture 
velocity of the bubble cap at bursting. 
Both features are uncontrolled in the field, and could explain the low correlation 
observed at Stromboli volcano between the amplitude of the acoustic wave and the vigor 
of the explosive event both in terms of mass of ejected fragments \cite{Marchetti09} 
and gas volume \cite{McGonigle09}.

%
%

{\small
{\bf Acknowledgments}
We thanks M. Ichihara for the fruitful discussions on a previous version of 
this manuscript. S. Lane is acknowledged for his useful comments, which improved 
the manuscript. D.L. was supported by FONDECYT Project \#1061253. 
V.V., D.L., J.C.G. and F. M. were supported by CNRS/CONICYT Project \#18640.
V.V. and J.C.G. thank the {\it Centre National de la Recherche Scientifique}
(CNRS, France) for supporting the research of their members in foreign laboratories.
}

%
%


\end{document}